\documentclass[prx,preprint,superscriptaddress]{revtex4-2}

\usepackage{amsfonts}
\usepackage{amsmath}
\usepackage{amssymb}
\usepackage{lineno}
\usepackage{color}
\usepackage{graphicx}%
\usepackage{makecell}
\usepackage{multirow}
\begin{document}


\title{Orbital-ordered ferromagnetic insulating state in tensile-strained SrCoO$_{3}$ thin films}

\author{Sheng-Chieh Huang}
\affiliation{Department of Physics, National Central University, Taoyuan City 320317, Taiwan}

\author{Kanchan Sarkar}
\affiliation{Universit\"at Ulm, Institut f\"ur Theoretische Chemie, Ulm 89081, Germany}
\affiliation{Department of Applied Physics and Applied Mathematics, Columbia University, New York, New York 10027, USA}

\author{Renata M. Wentzcovitch}
\affiliation{Department of Applied Physics and Applied Mathematics, Columbia University, New York, New York 10027, USA}
\affiliation{Department of Earth and Environmental Sciences, Columbia University, New York, New York 10027, USA}
\affiliation{Lamont–Doherty Earth Observatory, Columbia University, Palisades, New York 10964, USA}

\author{Han Hsu}
\email[Corresponding author: ]{hanhsu@ncu.edu.tw}
\affiliation{Department of Physics, National Central University, Taoyuan City 320317, Taiwan}

\date{\today}

\begin{abstract}

At ambient pressure, bulk SrCoO$_{3}$ is a ferromagnetic (FM) metal in cubic perovskite structure. By contrast, magnetic properties of epitaxial SrCoO$_{3}$ thin films, especially at high tensile strain ($\varepsilon \gtrsim 3$\%), remain unclear: Previous calculations had predicted antiferromagnetic (AFM) states more energetically favorable in this regime, but recent experiments indicated a FM insulating state. In this work, using first-principles calculations, we perform an extensive search for the structural, spin, magnetic, and orbital states of SrCoO$_{3}$ thin films. Our calculations indicate that at $0 < \varepsilon \lesssim 2.5$\%, SrCoO$_{3}$ favors a FM half-metallic state with intermediate-spin ($t_{2g}^{5}e_{g}^{1}$-like) Co exhibiting $d^{6}\underline L$ character. At $\varepsilon \gtrsim 2.5$\%, a FM insulating state with high-spin ($t_{2g}^{4}e_{g}^{2}$-like) Co dominates. This FM insulating state is achieved via complicated orbital ordering, cooperative Jahn--Teller distortion, and octahedral tilting about all three crystal axes.

\end{abstract}

\maketitle

\newpage

Perovskite oxides exhibit various properties of scientific and technological importance, including ferroelectricity, (anti)ferromagnetism, colossal magnetoresistance, and multiferroics. These properties, arising from the lattice, charge, spin, and orbital degrees of freedom in perovskite oxides, can be further engineered via epitaxial strains, offering this class of materials broad functionalities (see Refs.~\onlinecite{Rondinelli2011, Schlom2014, Lu2015, Damodaran2016} for reviews). Among diverse perovskite oxides, cobaltites are a unique family due to their tunable magnetic properties. For example, bulk LaCoO$_{3}$ is a nonmagnetic insulator with low-spin (LS, $t_{2g}^{6}e_{g}^{0}$) Co$^{3+}$ at low temperature ($T \lesssim 30$ K). Starting $\sim$30 K, bulk LaCoO$_{3}$ undergoes a thermally induced spin transition, becoming a paramagnetic insulator at $\sim$90 K. The detailed mechanism of this spin transition has been debated for decades but still remains unclear, as reviewed in Refs.~\onlinecite{Hsu2010, Raveau2012, Kobayashi2021, Watanabe2022}. By contrast, tensile-strained LaCoO$_{3}$ thin film is a ferromagnetic (FM) insulator at $T \lesssim 85$ K \cite{Fuchs2007, Fuchs2008, Freeland2008, Merz2010}. To explain this strain-induced FM insulating state, a mixture of high-spin (HS, $t_{2g}^{4}e_{g}^{2}$) and LS Co$^{3+}$ was proposed \cite{Hsu2012}, and this model is supported by various calculations and experiments \cite{Seo2012, Fujioka2015, Meng2018, Sterbinsky2018, Wang2019}.

\bigskip

In recent years, SrCoO$_{3-\delta}$ ($0 \leq \delta \leq 0.5$) thin films have also attracted significant attention due to their potential applications in catalysis and fuel cells and due to their composition-dependent chemical, structural, and magnetic properties \cite{Lee2011, Rivero2016, Choi2013, Tahini2016, Lu2020, Jeen2013NatMater, Lu2017, Jeen2013AdvMater, Gu2018, Hu2018, Callori2015, Hu2015, Petrie2016, Wang2020}. At ambient condition, bulk SrCoO$_{3}$ is a FM metal in cubic perovskite structure, and brownmillerite SrCoO$_{2.5}$ is an antiferromagnetic (AFM) insulator in orthorhombic structure \cite{Long2011}. Theory has shown that bulk SrCoO$_{3}$ contains intermediate-spin (IS, $t_{2g}^{5}e_{g}^{1}$-like) Co exhibiting $d^{6}\underline L$ character (nearly Co$^{3+}$ accompanied by O $2p$ holes) \cite{Potze1995, Zhuang1998, Abbate2002, Hoffmann2015, Lim2018, Hsu2018}, but magnetic properties of tensile-strained stoichiometric SrCoO$_{3}$ thin films remain controversial. Previous first-principles calculations had predicted AFM ordering energetically favorable at tensile strain $\varepsilon \gtrsim 2$\% \cite{Lee2011, Rivero2016}. Experiments, however, indicated that SrCoO$_{3}$ thin films remain FM metallic at $\varepsilon \approx 2$\% (grown on SrTiO$_{3}$ substrates) \cite{Jeen2013AdvMater, Gu2018, Hu2018, Petrie2016, Wang2020}. At $\varepsilon \approx 3$\% (grown on DyScO$_{3}$ substrates), while AFM ordering had been reported \cite{Callori2015}, other experiments indicated that the AFM ordering is induced by oxygen deficiency \cite{Petrie2016}. Furthermore, a recent work with high-quality stoichiometric SrCoO$_{3}$ thin films reported a FM insulating state at $\varepsilon \approx 3$\%~\cite{Wang2020}, in direct contradiction with theoretical predictions. In light of the HS+LS mixture in LaCoO$_{3}$ thin films, FM insulating state in SrCoO$_{3}$ thin films may also arise from spin transition and possible spin or orbital orderings. These factors, however, have long been ignored. An extensive search (via calculation) for the structural, magnetic, spin, and orbital states of SrCoO$_{3}$ thin films may thus clarify the above-mentioned controversy and further elucidate the magnetic properties and potential functionalities of SrCoO$_{3}$ thin films.

\bigskip

In this work, all calculations are performed using the {\sc Quantum ESPRESSO} (QE) codes \cite{PWscf2017}. Projected augmented wave (PAW) datasets are generated using the EPAW code \cite{EPAW2017, EPAW2018}, as detailed in Sec.~S1 of Supplemental Material (SM) \cite{SM}. The generalized-gradient approximation + Hubbard $U$ (GGA+$U$) method is adopted, with PBE-type GGA \cite{PBE}. Epitaxial thin films are modeled with strained bulks: Constraining the in-plane lattice vectors while relaxing the out-of-plan lattice vector and atomic positions. To balance the efficiency and accuracy of our extensive search, a reasonably chosen $U$-parameter is essential. Using the HP code (implemented in QE) based on linear response theory \cite{HP, Cococcioni2005, Timrov2021}, we compute the self-consistent $U$ and lattice constant ($a_{0}^{\text{GGA}+U}$) of bulk SrCoO$_{3}$. The obtained $U=6.5$~eV is adopted throughout this work, and $a_{0}^{\text{GGA}+U}=3.849$~\AA~is in good agreement with experiments ($a_{0}^{\text{exp}}=3.829$ \AA~\cite{Long2011}).

\bigskip

\begin{figure*}[pt]
\begin{center}
\includegraphics[
]{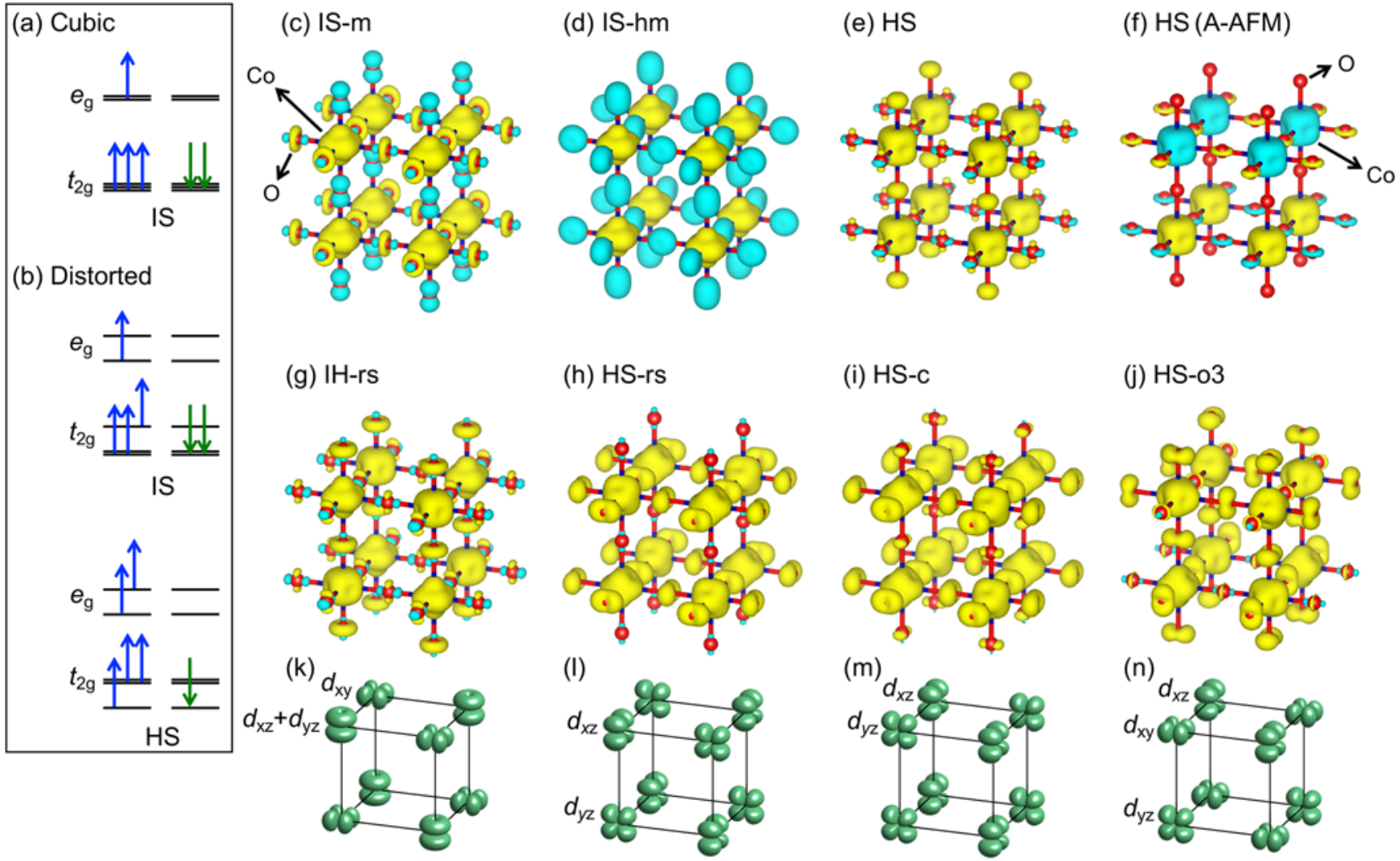}
\end{center}
\caption{(a,b) Schematic diagrams for the orbital occupations of Co$^{3+}$ in cubic and distorted crystal fields; (c--j) spin densities of the obtained spin/orbital states in the untilted ($a^{0}a^{0}c^{0}$) structures (see text for the nomenclatures), with isosurface values $\pm 0.02$~a.u.$^{-3}$ (yellow/cyan); (k--n) schematic diagrams for the minority-spin $3d$ electrons of the spin/orbital states in panels (g)--(j), respectively.}
\label{Fig:spinstates}
\end{figure*}

As mentioned above, bulk SrCoO$_{3}$ contains IS $d^{6}\underline L$ Co with two minority-spin electrons partially occupying the degenerate $t_{2g}$ orbitals [Fig.~\ref{Fig:spinstates}(a)]. With tensile strain,  HS ($t_{2g}^{4}e_{g}^{2}$-like) $d^{6}\underline L$ Co should also be considered. In distorted crystal fields, minority-spin electrons of IS and HS Co occupy two (e.g. $d_{xz}$ and $d_{yz}$) and one (e.g. $d_{xy}$) of the $t_{2g}$ orbitals, respectively, as schematically shown in Fig.~\ref{Fig:spinstates}(b). For transition-metal oxides, metal--insulator transition often coincides with charge, spin, and/or orbital orderings; these orderings should thus be considered for the possible FM insulating state in SrCoO$_{3}$ thin films. Nevertheless, interplay between the crystal structure, spin state, and magnetic/spin/orbital orderings are highly complicated. To better control these variables in our calculation, we begin with reduced degrees of freedom: Optimizing the Co-O bond lengths while keeping the CoO$_{6}$ octahedra untilted (Glazer notation $a^{0}a^{0}c^{0}$). To obtain spin or orbital orderings, cooperative Jahn--Teller (JT) distortions are manually imposed, followed by structural optimization. Within $a^{0}a^{0}c^{0}$, both IS and HS Co can be stabilized; various spin and orbital states are obtained, in FM and AFM ($A$, $C$, and $G$ types) orderings, as described below.

\bigskip

\begin{figure*}[pt]
\begin{center}
\includegraphics[
]{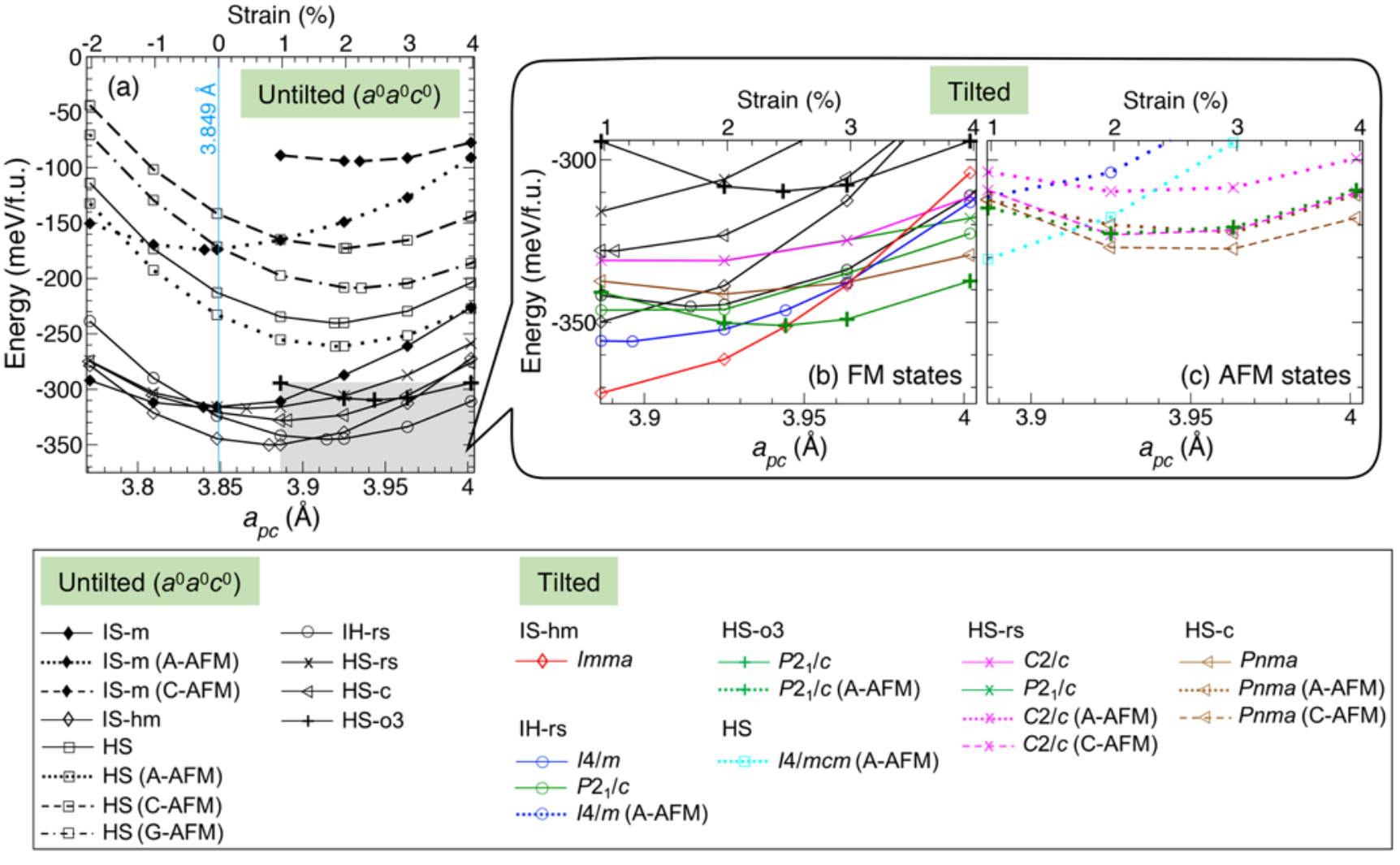}
\end{center}
\caption{Total energies of the obtained spin/orbital states at various strains (or equivalently, $a_{pc}$) in (a) the untilted and (b,c) tilted structures; for the latter, FM and AFM states are plotted in panels (b) and (c), respectively. The equilibrium energy and lattice constant of bulk SrCoO$_{3}$ are used as the reference for energy and strain (see text).}
\label{Fig:energy}
\end{figure*}

In Figs.~\ref{Fig:spinstates}(c)--\ref{Fig:spinstates}(j), spin densities $s(\bold r) \equiv n_{\uparrow}(\bold r)-n_{\downarrow}(\bold r)$ of the obtained states are plotted [$n_{\uparrow/ \downarrow}(\bold r)$: spin-up/down electron density]. For IS Co, FM ordering is favored; a metallic (IS-m) state [Fig.~\ref{Fig:spinstates}(c)] and a more energetically favorable half-metallic (IS-hm) state [Fig.~\ref{Fig:spinstates}(d)] are obtained (see Fig.~\ref{Fig:energy} for their energies). For both states, spin densities at the Co sites exhibit $e_{g}+d_{xy}$ character, indicating two spin-down electrons occupying the $d_{xz}$ and $d_{yz}$ orbitals. The main difference between these two states is that IS-hm has more prominent local magnetic moments at the O sites, resulting from O $2p$ holes. For HS Co, without orbital ordering, FM ordering [Fig.~\ref{Fig:spinstates}(e)] is \textit{not} favored; $A$-type AFM ($A$-AFM) [Fig.~\ref{Fig:spinstates}(f)] is (see also Fig.~\ref{Fig:energy}). For both magnetic orderings, spin densities at the Co sites exhibit prominent dimples with $d_{xy}$ character, indicating one minority-spin electron occupying the $d_{xy}$ orbital. This type of HS Co is thus referred to as HS-$d_{xy}$ Co. (Likewise, the minority-spin electron of HS-$d_{xz}$/$d_{yz}$ Co occupies the $d_{xz}$/$d_{yz}$ orbital). In addition to pure IS and HS states, one spin-ordered and three orbital-ordered states are obtained, as shown in Figs.~\ref{Fig:spinstates}(g)--\ref{Fig:spinstates}(j), respectively: A mixture of IS+HS-$d_{xy}$ Co in rock-salt ordering (IH-rs), a mixture of HS-$d_{xz}$+HS-$d_{yz}$ Co in rock-salt (HS-rs) and columnar orderings (HS-c), and a mixture of all three types of HS Co in a more complicated ordering (HS-o3): HS-$d_{xy}+$HS-$d_{yz}$ checkerboards inter-layered with HS-$d_{xy}+$HS-$d_{xz}$ checkerboards. To better visualize the spin/orbital orderings of these states, their spin-down $3d$ electrons are schematically plotted in Figs.~\ref{Fig:spinstates}(k)--\ref{Fig:spinstates}(n), respectively. For these four states, FM ordering is favored (see Fig.~\ref{Fig:energy} and Table~SIV in SM \cite{SM}). 

\bigskip

In Fig.~\ref{Fig:energy}, total energies of the obtained states at various strains (or equivalently, in-plane pseudocubic lattice parameter $a_{pc}$) are plotted, where symbols indicate spin/orbital states, line formats indicate magnetic orderings, and line colors indicate structures. The equilibrium energy and lattice constant of cubic perovskite SrCoO$_{3}$ are used as the reference for energy ($0$ eV) and epitaxial strain $\varepsilon \equiv a_{pc}/a_{0}^{\text{GGA}+U}-1$, respectively. By analyzing the results of the untilted ($a^{0}a^{0}c^{0}$) structures [Fig.~\ref{Fig:energy}(a)], effects of Co spin state and magnetic/orbital orderings can be deduced. Overall, IS Co favors FM ordering, and the IS-hm state (empty diamond, solid line) is the most energetically favorable state, with energy minimum $E_{min}^{\text{IS-hm}}=-350$~meV/f.u. at $a_{pc}=3.880$~\AA. Energy of the IS-m state (filled diamond, solid line) is higher than IS-hm at $0 < \varepsilon \leq 4$\% and is even higher in $A$-AFM (filled diamond, dotted line), $C$-AFM (filled diamond, dashed line), and $G$-AFM orderings (beyond the graph range). By contrast, for HS Co without orbital ordering, $A$-AFM ordering (square, dotted line) is more favorable ($E_{min}^{\text{HS (A-AFM)}}=-261$~meV/f.u. at $a_{pc}=3.920$~\AA) than FM (square, solid line). Nevertheless, HS $A$-AFM is still not energetically competitive at $\varepsilon > 1$\%; the spin/orbital-ordered states (IH-rs, HS-rs, HS-c, and HS-o3) are, as highlighted by the shaded area. All these spin/orbital-ordered states favor FM ordering. When in AFM orderings, their energies increase by $25$--$215$~meV/f.u. (see Table~SIV in SM \cite{SM} for $E_{min}$'s of all the obtained states in all magnetic orderings). For clarity, we only plot the FM results in Fig.~\ref{Fig:energy}(a).
 
\bigskip

To search for the most energetically favorable structure for each spin/orbital state, octahedral tilting must be considered. Given that SrCoO$_{3}$ thin films are often grown on substrates with equivalent in-plane crystal axes, we require the same type of octahedral tilts about the in-plane crystal axes: both in-phase (+), both out-of-phase (--), or both untilted (0). Combined with the tilting about the out-of-plane axis, nine types of tilts (including $a^{0}a^{0}c^{0}$) are considered, as tabulated in Table~I, along with the associated space groups. (\textit{Note}: Based on group-theoretical analysis, some of the tabulated space groups are associated with less restricted tilts, including: $Immm$, $Pnnn$, and $C2/m$ associated with $a^{+}b^{+}c^{+}$ instead of $a^{+}a^{+}c^{+}$, $C2/c$ associated with $a^{-}a^{-}c^{-}$ instead of $a^{-}a^{-}c^{0}$, and $P\bar1$ associated with $a^{-}b^{-}c^{-}$ instead of $a^{-}a^{-}c^{-}$ \cite{Howard1998, Howard2003, Carpenter2009}. To our best knowledge, no group-theoretical analysis is available for the structures of HS-o3.) For each spin/orbital state, octahedral tilts listed in Table~I are manually imposed, followed by structural optimization. Energies of the most favorable structures for FM and AFM orderings are plotted in Figs.~\ref{Fig:energy}(b) and \ref{Fig:energy}(c), respectively. Overall, even with octahedral tilting, AFM orderings are still unfavorable, despite that HS $A$-AFM becomes quite competitive in the $I4/mcm$ structure [cyan square, Fig.~\ref{Fig:energy}(c)]. Effects of octahedral tilting on magnetic ordering are basically insignificant. In Fig.~\ref{Fig:energy}(b), some states in the untilted structures are included for references (black symbols). With octahedral tilting, the IS-hm ($Imma$) and HS-o3 ($P2_{1}/c$) states, indicated by red diamond and green plus, respectively, are the most energetically favorable state at $\varepsilon \lesssim 2.5$\% and $\varepsilon \gtrsim 2.5$\%, respectively. Their energies are lower than their untilted counterparts in $P4/mmm$ (black diamond) and $P4_{2}/mnm$ symmetries (black plus) by $\sim$$20$ and $\sim$$50$ meV/f.u., respectively. 



\bigskip

\begin{figure}[pt]
\begin{center}
\includegraphics[
]{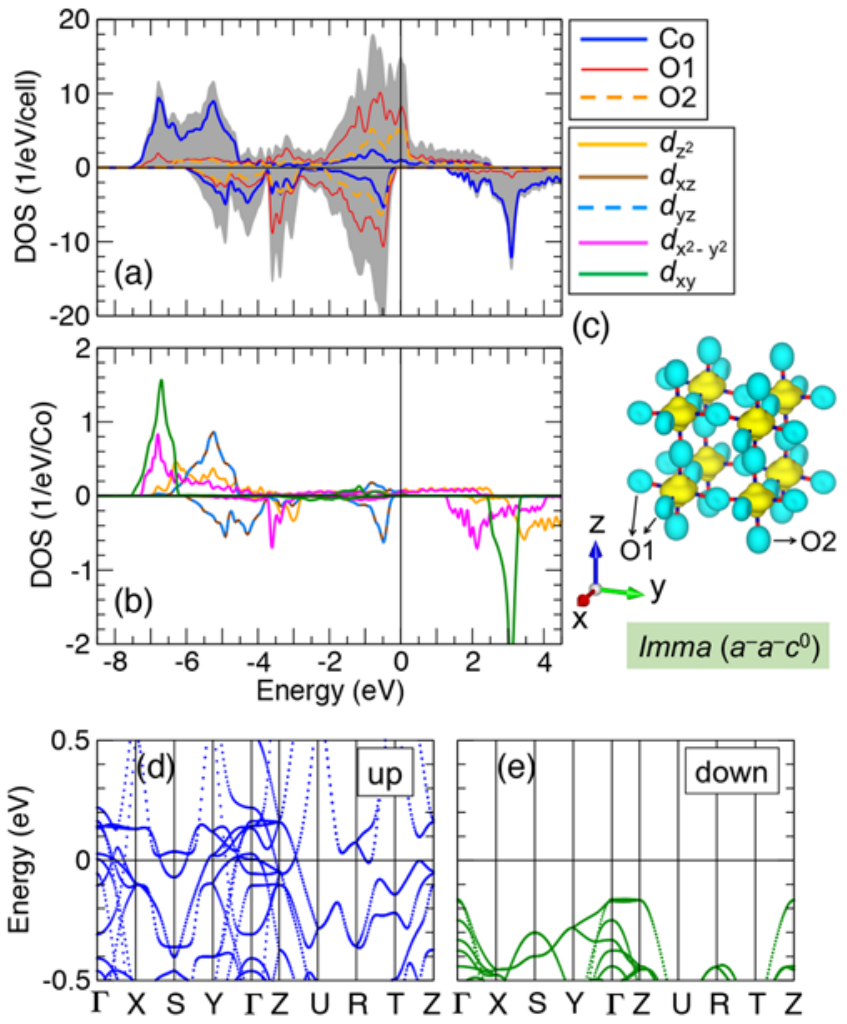}
\end{center}
\caption{Electronic structure of the IS-hm ($Imma$) state at $a_{pc}=3.963$ \AA~($\varepsilon = 3$\%). (a) Total (shade) and projected DOS onto each atomic species (lines); (b) projected DOS onto the Co $3d$ orbitals; (c) spin density; (d,e) band structure of the spin-up/down channel.}
\label{Fig:dos-IS}
\end{figure}

\begin{figure}[pt]
\begin{center}
\includegraphics[
]{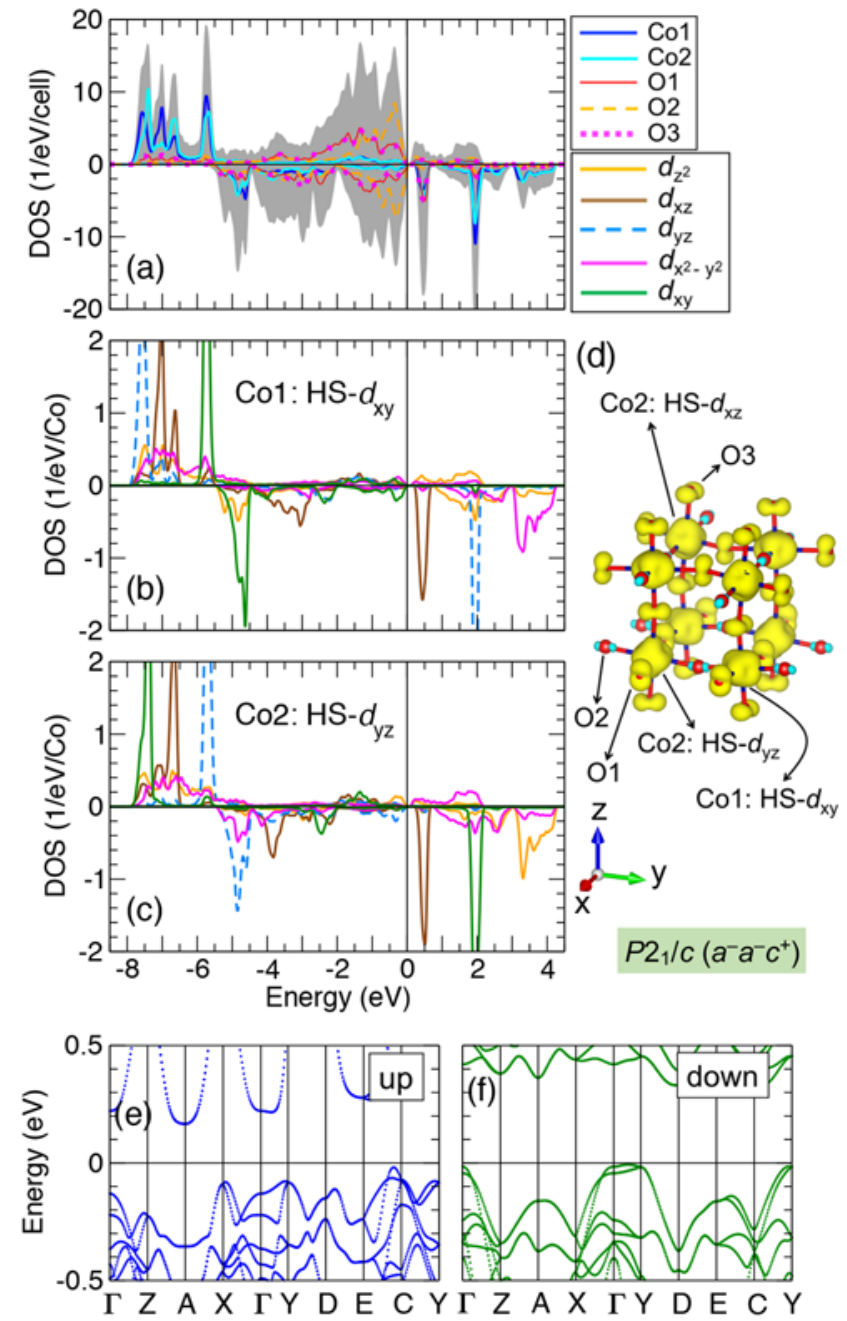}
\end{center}
\caption{Electronic structure of the HS-o3 ($P2_{1}/c$) state at $a_{pc}=3.963$ \AA~($\varepsilon = 3$\%). (a) Total (shade) and projected DOS onto each atomic species (lines); (b,c) projected DOS onto the $3d$ orbitals of Co1 and Co2 (see text); (d) spin density; (e,f) band structure of the spin-up/down channel.}
\label{Fig:dos-HS}
\end{figure}

In Figs.~\ref{Fig:dos-IS} and \ref{Fig:dos-HS}, density of states (DOS), band structure, and spin density of the IS-hm ($Imma$) and HS-o3 ($P2_{1}/c$) states at $a_{pc}=3.963$~\AA~($\varepsilon=3$\%) are plotted. Expectedly, their spin densities [Figs.~\ref{Fig:dos-IS}(c) and \ref{Fig:dos-HS}(d)] show great resemblance to their $a^{0}a^{0}c^{0}$ counterparts [Figs.~\ref{Fig:spinstates}(d) and \ref{Fig:spinstates}(j)]. For the IS-hm ($Imma$) state, $d_{xz}$ and $d_{yz}$ orbitals are equivalent, and their DOS curves coincide [Fig.~\ref{Fig:dos-IS}(b)]. Their occupation numbers in the spin-down channel are both $0.998$, and the local magnetic moment at the Co site $\mu_{\text{Co}}=2.078\mu_{B}$. Within the $Imma$ structure, there are two inequivalent O sites, O1 and O2 [Fig.~\ref{Fig:dos-IS}(c)], with local magnetic moments $\mu_{\text{O1}}=-0.303\mu_{B}$ and $\mu_{\text{O2}}=-0.504\mu_{B}$. For the HS-o3 ($P2_{1}/c$) state, HS-$d_{yz}$ and HS-$d_{xz}$ Co are crystallographically equivalent; there are thus only two Co sites: Co1 (HS-$d_{xy}$) and Co2 (HS-$d_{yz}$/HS-$d_{xz}$). Here we plot the projected DOS onto the $3d$ orbitals of HS-$d_{xy}$ and HS-$d_{yz}$ Co in Figs.~\ref{Fig:dos-HS}(b) and \ref{Fig:dos-HS}(c) (see Fig.~S1 in SM \cite{SM} for HS-$d_{xz}$ Co). Clearly, they each have one spin-down electron, occupying the $d_{xy}$ and $d_{yz}$ orbital, respectively; their local magnetic moments $\mu_{\text{Co1}}=2.328\mu_{B}$ and $\mu_{\text{Co2}}=2.281\mu_{B}$. In the $P2_{1}/c$ structure, there are three O sites [Fig.~\ref{Fig:dos-HS}(d)]. Among them, O1 and O3 have nearly the same DOS curves [Fig.~\ref{Fig:dos-HS}(a)] and significant local magnetic moments ($\mu_{\text{O1}}=0.337\mu_{B}$ and $\mu_{\text{O3}}=0.336\mu_{B}$), while O2 is nearly nonmagnetic ($\mu_{\text{O2}}=-0.032\mu_{B}$). Evident from Figs.~\ref{Fig:dos-HS}(e) and \ref{Fig:dos-HS}(f), an energy gap of $0.188$~eV is opened. Remarkably, only in the $P2_{1}/c$ structure is the HS-o3 state insulating; all other structures are metallic (see Table~SV and Fig.~S2 in SM \cite{SM}). Furthermore, the HS-rs and HS-c states are also metallic in all structures (see Tables~SVI and SVII in SM \cite{SM}). It can thus be concluded that the FM insulating state in tensile-strained SrCoO$_{3}$ is achieved via complicated orbital ordering, cooperative JT distortion, and octahedral tilting about all three axes. Despite its monoclinic structure, the out-of-plane lattice vector of the HS-o3 ($P2_{1}/c$) state is almost perpendicular to the (001) plane, consistent with the high-resolution transmission electron microscopy (HRTEM) image obtained in experiments \cite{Wang2020} (see Fig.~S3 in SM \cite{SM}).

\bigskip

Comparing with previous calculations for SrCoO$_{3}$ thin films (e.g. Refs.~\onlinecite{Lee2011, Rivero2016}), the major difference of this work includes a more rigorously determined $U$-parameter, investigations for spin/orbital orderings, and a systematic structure search for each spin/orbital state. In Ref.~\onlinecite{Lee2011}, a fairly small $U=2.5$ and $J=1.0$~eV were adopted; the ground state of SrCoO$_{3}$ was found to be FM cubic perovskite, and AFM states with cation displacements ($Amm2$ and $P4mm$ structures) were found favorable at $\varepsilon \gtrsim 2$\%. With a larger $U$, SrCoO$_{3}$ favors distorted structures than cubic by up to $> 300$~meV/f.u., as shown in Ref.~\onlinecite{Rivero2016} ($U=6.0$~eV) and this work, suggesting that cubic perovskite may be a high-temperature phase of SrCoO$_{3}$. Also reported in Ref.~\onlinecite{Rivero2016} is that for $U=6.0$~eV, structures with cation displacements are highly unfavorable (by hundreds of meV/f.u.). We therefore do not consider cation displacements in this work. Remarkably, the ground state reported in Ref.~\onlinecite{Rivero2016} is the same as our HS-c state in the $P4/mbm$ (untilted) structure, indicated as black triangles in Fig.~\ref{Fig:energy}(b) (see Fig.~S4 in SM \cite{SM} for a detailed complarison). Nevertheless, spin/orbital orderings were not addressed in these previous works.

\bigskip

In summary, we perform an extensive search for the structural, spin, magnetic, and orbital states of tensile-strained SrCoO$_{3}$ thin films via GGA+$U$ calculations. Our results indicate that at tensile strain $\varepsilon \lesssim 2.5$\%, SrCoO$_{3}$ favors an IS half-metallic ($Imma$) state with $d^{6}\underline L$ character. At $\varepsilon \approx 2.5$\%, a simultaneous structural, spin, orbital, and metal--insulator transition occurs; a FM insulating ($P2_{1}/c$) state with HS Co in a complicated orbital ordering emerges. At $\varepsilon \gtrsim 2.5$\%, SrCoO$_{3}$ favors this FM insulating state, consistent with recent experiments. The energy gap of this state is opened via complicated orbital ordering, cooperative JT distortion, and octahedral tilting about all three crystal axes. 

\bigskip
 
\textbf{Acknowledgments} This work was primarily supported by the National Science and Technology Council (formerly Ministry of Science and Technology) of Taiwan under Grants No.~MOST 107-2119-M-009-009-MY3, 111-2124-M-A49-009, 107-2112-M-008-022-MY3, 110-2112-M-008-033, and 111-2112-M-008-032. R.M.W. acknowledges support from DOE grant DE-SC0019759 and NSF grant EAR-1819126. We thank National Center for High-performance Computing (NCHC) of Taiwan for the computation resources.

\newpage
\begin{table*} 
\caption {Imposed octahedral tilts and associated space groups (and lattice systems) for all spin/orbital states. T: tetragonal; O: orthorhombic; M: monoclinic; Tri: triclinic. \textit{Note}: Some space groups are associated with less restricted tilts (see text).}
\begin{ruledtabular}
\begin{tabular}{c c c c c c}
\makecell {Imposed tilts} &   \makecell {IS-m/hm or HS}   &    IH-rs    &    \makecell {HS-rs}    
&   \makecell {HS-c}    &    \makecell{HS-o3} \\ 
\hline
$a^{0}a^{0}c^{0}$ & $P4/mmm$ (T) & $I4/mmm$ (T) & $I4/mcm$ (T) & $P4/mbm$ (T) & $P4_{2}/mnm$ (T)   \\
\hline
$a^{0}a^{0}c^{+}$ & $P4/mbm$ (T) & $P4/mnc$ (T) & $P4_{2}/mbc$ (T) & $Pbam$ (O) & $Pnnm$ (O)   \\
\hline
$a^{0}a^{0}c^{-}$ & $I4/mcm$ (T) & $I4/m$ (T) & $Ibam$ (O) & $P4_{2}/mbc$ (T) & $P4_{2}/m$ (T)   \\
\hline
$a^{+}a^{+}c^{0}$ & $I4/mmm$ (T) & $P4_{2}/nnm$ (T) & $P4/nnc$ (T) & $I4/m$ (T) & $P4_{2}/n$ (T)   \\
\hline
$a^{+}a^{+}c^{+}$ & $Immm$ (O) & $Pnnn$ (O) & $Pnnn$ (O)  & 
$C2/m$ (M) & $P2/c$ (M)   \\
\hline
$a^{+}a^{+}c^{-}$ & $P4_{2}/nmc$ (T) & $P4_{2}/n$ (T) & $Ccca$ (O) & $P4_{2}/n$ (T) & $P4_{2}/n$ (T)   \\
\hline
$a^{-}a^{-}c^{0}$ & $Imma$ (O) & $C2/m$ (M) & $C2/c$ (M) & $Pnma$ (O) & $P2_{1}/c$ (M)   \\
\hline
$a^{-}a^{-}c^{+}$ & $Pnma$ (O) & $P2_{1}/c$ (M) & $P2_{1}/c$ (M) & $Pnma$ (O) & $P2_{1}/c$ (M)   \\
\hline
$a^{-}a^{-}c^{-}$ & $C2/c$ (M) & $P\bar{1}$ (Tri) & $C2/c$ (M) & $P2_{1}/c$ (M) & $P\bar{1}$ (Tri) 
\end{tabular}
\end{ruledtabular}
\end{table*}

\newpage

\newpage

\newpage

\newpage


\end{document}